\RequirePackage{arydshln}
\documentclass[aps,twocolumn,nofootinbib,superscriptaddress,preprintnumbers,pra,10pt]{revtex4-1}

\usepackage{float}
\usepackage{colortbl}
\usepackage{amsmath,amssymb,amsfonts, bm,bbm,slashed}
\usepackage{dsfont} 
\usepackage{hyperref}
\usepackage{graphicx}
\usepackage{enumitem}
\usepackage{arydshln}
\usepackage{mathtools}
\usepackage{bbold}

\makeatletter
\g@addto@macro\bfseries{\boldmath}
\makeatother

\newcommand{\be} {\begin{equation}}
\newcommand{\ee} {\end{equation}}
\newcommand{\bea} {\begin{eqnarray}}
\newcommand{\eea} {\end{eqnarray}}
\newcommand{\no} {\nonumber}
\newcommand{\ba} {\begin{array}}
\newcommand{\ea} {\end{array}}
\newcommand{\gsim}{\lower.7ex\hbox{$\;\stackrel{\textstyle>}{\sim}\;$}}
\newcommand{\lsim}{\lower.7ex\hbox{$\;\stackrel{\textstyle<}{\sim}\;$}}
\renewcommand{\Re}{{\rm Re}}

\newcommand{\cC}{\mathcal{C}}
\newcommand{\cO}{\mathcal{O}}
\newcommand{\cA}{\mathcal{A}}
\newcommand{\cB}{\mathcal{B}}
\newcommand{\xZ}{x_{\Zp}}
\newcommand{\xG}{x_{G^\prime}}
\newcommand{\mU}{m_U}
\newcommand{\mZ}{m_{Z^\prime}}
\newcommand{\mG}{m_{G^\prime}}
\newcommand{\Gp}{G^\prime}
\newcommand{\Zp}{Z^\prime}

\begin{document}

\preprint{ZU-TH-45/19}

\title{Vector Leptoquarks Beyond Tree Level}
 
\author{Javier Fuentes-Mart\'{\i}n}
\email{fuentes@physik.uzh.ch}
\affiliation{Physik-Institut, Universit\"at Zu\"rich, CH-8057 Z\"urich, Switzerland}
\author{Gino Isidori}
\email{isidori@physik.uzh.ch}
\affiliation{Physik-Institut, Universit\"at Zu\"rich, CH-8057 Z\"urich, Switzerland}
\author{Matthias K\"{o}nig}
\email{matthias.koenig@uzh.ch}
\affiliation{Physik-Institut, Universit\"at Zu\"rich, CH-8057 Z\"urich, Switzerland}
\author{Nud{\v z}eim Selimovi{\'c}}
\email{nudzeim@physik.uzh.ch}
\affiliation{Physik-Institut, Universit\"at Zu\"rich, CH-8057 Z\"urich, Switzerland}

\begin{abstract}
\vspace{5mm}
Models with massive vector  leptoquarks, resulting from an $SU(4)$ 
gauge symmetry spontaneously broken at the TeV scale, 
are of great phenomenological interest given the current  ``anomalies"
in semileptonic $B$ decays. 
We analyze the relations between low- and high-energy observables 
in such class of models to next-to-leading order accuracy in the 
$SU(4)$ gauge coupling $g_4$.  For large values of $g_4$, motivated by recent $B$-physics data,
one-loop corrections are sizeable.
The main 
effect is an enhanced contribution at low-energy, at fixed on-shell couplings. This result has 
important implications for current and future high-energy searches of vector  leptoquark
models.
% related to the $B$-physics anomalies.
\vspace{3mm}
\end{abstract}

\maketitle

\allowdisplaybreaks

%%%%%%%%%%%%%%%%%%%%%%%%%%%%%%%%%%%%%%%%%%%%%%%%%%
\section{Introduction}\label{sec:intro}
%%%%%%%%%%%%%%%%%%%%%%%%%%%%%%%%%%%%%%%%%%%%%%%%%%

A natural expectation of grand-unified theories, where a single fermion representation contains both 
quark and lepton fields, is the presence of massive vector leptoquarks, i.e.~vector fields transforming quarks 
into leptons and vice versa. 
 One of the most appealing  constructions of this type is the model 
proposed by Pati and Salam (PS)~\cite{Pati:1974yy}, where quarks and leptons are unified 
in fundamental representations of the  $SU(4)$ gauge group. The breaking $SU(4) \to  SU(3)_c \times U(1)$  gives rise to a single
vector leptoquark, $U_1$, transforming as $(\mathbf{3},\mathbf{1},2/3)$ under the Standard Model (SM) gauge symmetry.

A renewed phenomenological interest in the PS model has been triggered by the recent 
$B$-phsiycs anomalies, i.e.~the hints of Lepton Flavor Universality (LFU) violations in semi-leptonic $B$ 
decays~\cite{Lees:2013uzd,Aaij:2015yra,Hirose:2016wfn,Aaij:2017deq,Aaij:2014ora,Aaij:2017vbb}.
Already in the early phenomenological attempts to explain these anomalies~\cite{Alonso:2015sja, Barbieri:2015yvd, Calibbi:2015kma,Buttazzo:2017ixm},
it appeared that a TeV-scale $U_1\sim(\mathbf{3},\mathbf{1},2/3)$ field, coupled mainly to the third generation,
is an excellent mediator to account for all available data.

The problem of the original PS model in this context is the flavor-universal nature of the $U_1$, 
which has to be very heavy in order to satisfy the tight bounds derived from its coupling to light SM fermions.
This problem can be overcome in a natural way with two main ingredients: enlarging the gauge group~\cite{DiLuzio:2017vat},
and allowing gauge non-universal charges to the SM fermions~\cite{Bordone:2017bld}.
These two ingredients have been been analyzed in a series of recent papers~\cite{DiLuzio:2017vat,Bordone:2017bld,Greljo:2018tuh,DiLuzio:2018zxy,Cornella:2019hct}.
The proposed models have a few differences, but the TeV-scale dynamics is always characterized by the 
gauge group  $SU(4) \times  SU(3) \times SU(2) \times U(1)$,
effectively acting in a family non-universal way. These models, which we 
collectively denote as \emph{4321 models}, provide both a successful ultraviolet (UV) completion
for effective descriptions of the $B$-physics anomalies and, at the same time, 
represent a first step to shed light on the origin of SM mass hierarchies~\cite{Bordone:2017bld}
(alternatives approaches to embed the $U_1$ in extended PS-type models have been proposed in~\cite{Assad:2017iib,Calibbi:2017qbu}).

The spontaneous symmetry breaking of  the 4321 gauge group to  $G_{\rm SM} = SU(3)_c \times SU(2)_L\times U(1)_Y$
 gives rise to two additional 
massive vectors beside  the $U_1$: a color octet $\Gp\sim(\mathbf{8},\mathbf{1},0)$, commonly referred to as coloron, and a color singlet $\Zp\sim(\mathbf{1},\mathbf{1},0)$. 
As pointed out in~\cite{Baker:2019sli}, the presence of (at least) these additional states is a general feature of any 
UV completion of a flavor non-universal $U_1$. These states are indeed present also in UV 
completions based on new strongly interacting 
dynamics~\cite{Barbieri:2017tuq, Blanke:2018sro}.

So far, the dynamics of these heavy vectors has been analyzed only at leading order in the leptoquark (LQ) coupling. 
Next-to-leading order (NLO) effects in QCD have been studied, both at low energies~\cite{Aebischer:2018acj} (in the corrections to the coefficients of the corresponding four-fermion operators), and at high energies~\cite{Hammett:2015sea} (in LQ production and decay at colliders). However, NLO corrections associated to the heavy dynamics have never been analyzed.
In the absence of a UV completion, neglecting these corrections is a necessary choice.
But the validity of this approximation is questionable given that the coupling of the leptoquark to SM fermions 
must be large ($2 \lsim g_4 \lsim 3$)
in order to explain $B$-physics data, while being consistent with collider searches.

Employing a simplified 4321 model, which provides a consistent and sufficiently general description of
the heavy-vector dynamics, we present, for the first time, an estimate of the
NLO corrections associated to the leptoquark coupling ($g_4$).
Since the latter is large, we work in the limit where all SM couplings (both gauge and Yukawa) 
are set to zero. This limit simplifies the calculation, and isolates all the leading effects proportional to $\alpha_4 = g_4^2/(4\pi)$, without loss 
of generality. The results obtained this way
are applicable to all the realistic 4321 models proposed in the literature. 
Being interested only in the physical effects generated by these quantum corrections, 
%and particularly in the modified relations between low- and high-energy observables,
we adopt an on-shell renormalization scheme: masses and couplings of the heavy states 
are defined from their on-shell production and decay processes, 
and we evaluate NLO corrections to low-energy amplitudes in terms of these parameters.

%%%%%%%%%%%%%%%%%%%%%%%%%%%%%%%%%%%%%%%%%%%%%%%%%%
\section{The model}\label{sec:model}
%%%%%%%%%%%%%%%%%%%%%%%%%%%%%%%%%%%%%%%%%%%%%%%%%%

We consider a simplified version of the 4321~model, where we set the SM gauge couplings to zero,
$g_{1,2,3}=0$.  Furthermore, we ignore the SM Higgs sector, meaning that the model has
an exact $SU(3)_{12} \times SU(2)_L \times U(1)_Y$ global symmetry: the $SU(3)_{12}$ group acts only on the 
light generations, which decouple being $SU(4)$ singlets. The $SU(2)_L \times U(1)_Y$ group 
is flavor universal. The only non-trivial dynamics is that of the $SU(4)$ gauge group,
with coupling $g_4$.
% controls both the dynamics of the heavy vectors,  % and their couplings to the SM fermions. 

The non-decoupling fermion fields are one $SU(2)_L$ doublet, $\psi_L$, and two $SU(2)_L$  singlets, $\psi_u$ and $\psi_d$.
As we discuss later, these fields can be identified with the SM third generation, up to (small) mixings with the light families and/or
mixings with heavy exotic fermions. In the SM-gaugeless limit, these massless fields consist of four identical chiral fermions transforming in the fundamental representation of $SU(4)$.

The spontaneous breaking of $SU(4)$ is achieved by two $SU(4)$-fundamental scalars, $\Omega_1$ and  $\Omega_3$, transforming as singlet and triplet under $SU(3)$, respectively.
The Lagrangian of this simplified model reads
\bea
\mathcal{L} &=&-\frac{1}{4}H^\alpha_{\mu\nu}H^{\alpha\,\mu\nu}+\sum_{i=1,3} \,(D_\mu\Omega_i)^\dagger\,D_\mu\Omega_i   \no\\
&& + \sum_{f=L,u,d}  i \bar\psi_f \slashed{D}\psi_f +V(\Omega_i)\,,
\eea
where $H_{\mu\nu}^\alpha$ ($\alpha=1,\dots,15$) is the $SU(4)$ field-strength tensor.  We further assume that all the radial modes are much heavier than the vector resonances
($M^2_R \gg g^2_4 v^2_{3(1)})$,
with $v_{1(3)}$ denoting the vacuum expectation of $\Omega_1(\Omega_3)$. This way, we can restrict the attention to the dynamics of 
gauge fields, Goldstone bosons, and fermions.\footnote{An extended analysis including radial modes and 
heavy fermions will be presented elsewhere~\cite{Long}.}

After spontaneous symmetry breaking,
no physical scalars remain massless and all $SU(4)$ gauge fields acquire a mass.
The latter can be identified with the massive vector resonances of the realistic $4321$ 
models~\cite{DiLuzio:2017vat,Bordone:2017bld,Greljo:2018tuh,DiLuzio:2018zxy,Cornella:2019hct}.
The charge and mass eigenstates of the $SU(4)$ gauge bosons $H^\alpha$ are 
\bea
&& \Gp_\mu=H^a_\mu\,,\qquad  \Zp_\mu=H_\mu^{15}\,,  \no\\
&& U_\mu^{1,2,3}=\frac{1}{\sqrt{2}}\left(H_\mu^{9,11,13}-iH_\mu^{10,12,14}\right)\,,
\eea
with masses $\mG^2=(g^2_4/2) v^2_3$,  $m^2_U= (g^2_4/4) (v_1^2+v_3^2)$ 
and $\mZ^2=  (3 g^2_4/8) (v_1^2+ v_3^2/3)$.
In the limit $v_1=v_3$ there is a residual custodial $SU(4)$ global symmetry 
and all massive vectors are degenerate.

In the mass eigenbasis, the interactions between vectors and fermions read
\bea
\mathcal{L}_{\rm int} &\supset&    \frac{g_4}{\sqrt{2}}\, \left[ U_\mu\,\bar\psi_q\,\gamma^\mu\psi_\ell  +{\rm h.c.} \right]
+ g_4\, \Gp_\mu \,\bar\psi_q\gamma^\mu\,T^a\psi_q     \no\\
&& +  \frac{g_4 \sqrt{6}}{4}\, \Zp_\mu \,(\bar \psi\, T_{\rm B-L}  \gamma^\mu\, \psi)~, \label{eq:U1Lag}
\eea
where $\psi = (\psi_q\;\psi_\ell)^\intercal$ are $SU(4)$ fermion multiplets and $T_{\rm B-L}={\rm diag}(\tfrac{1}{3},\tfrac13,\tfrac13,-1)$.

%%%%%%%%%%%%%%%%%%%%%%%%%%%%%%%%%%%%%%%%%%%%%%%%%%
\section{One-loop results}\label{sec:nlo}
%%%%%%%%%%%%%%%%%%%%%%%%%%%%%%%%%%%%%%%%%%%%%%%%%%

Our simplified model is completely renormalizable only after the inclusion of the radial modes; 
however, this does not prevent us from obtaining finite and gauge-invariant results in the on-shell scheme, 
once we add an appropriate set of counterterms (as in the non-linear sigma model). 
The results obtained this way are correct up to finite terms of $O(m_V^2/M_R^2)$ which we assume to be small.
The explicit inclusion of the Goldstone modes ensure gauge-invariant results.
All partial results reported below are obtained in the Feynman gauge. 

\subsection{Vertex corrections}
We start analyzing the correction to the three-point functions 
with one external heavy vector and two light fermions. The modified LQ vertex function assumes the form 
\be
\mathcal{A}^U_\mathrm{vertex}  =  i \frac{g_4}{\sqrt{2} }  \epsilon_\mu(q) \psi_q \gamma^\mu \psi_\ell \times \left[1 +  \frac{\alpha_4}{4\pi}  \delta V_U(s) \right]\,,
\ee
where $\epsilon_\mu(q)$ is the LQ polarization vector and $s=q^2$.
Using dimensional regularization in $d=4-2\epsilon$, we find:
\be
\delta V_U(s)  =  \frac{47}{8}\left(\frac{1}{\epsilon}+\log\frac{\mu^2}{m_U^2} \right) + \Lambda^0_U 
+\Lambda_U(s, \{m^2_{V_i}\})~, 
\label{eq:Vos}
\ee
where $\Lambda^0_U$ is constant and the $s$-dependent term, satisfying $\Lambda_U(0, \{m^2_{V_i}\}) =0$, can be expressed as
\bea
\Lambda_U(s, \{m^2_{V_i}\})  &=& -\frac{1}{8} \Lambda_2(s, \mZ) + 2 \Lambda_4(s, \mZ, \mU) \no\\
&& +4 \Lambda_4(s, \mG, \mU)\,,
\eea
in terms of the loop function reported in~\cite{Bohm:1986rj}.
Note that the coefficient of the UV divergence is nothing but $C_A+C_F$.
After renormalization, defining the renormalized coupling from the on-shell LQ  vertex, 
the finite vertex correction (for off-shell processes) reads 
\be
  \delta V_U(s)_r =
 \delta V_U(s) - \Re\left[ \delta V_U(\mU^2) \right]~.
\ee
Under these renormalization conditions the
constant terms in (\ref{eq:Vos}) do not play a role in physical observables.
At $s=0$, and in the $SU(4)$ custodial limit for the vector-boson masses, we find
\be
  \delta V_U(0)^{(\rm{cust.})}_r =  -\frac{27}{16}+ \frac{17}{12} \pi^2-2\sqrt{3}\pi \approx  1.41~.
  \label{eq:dVcust}
\ee 
 
Proceeding in a similar way for the coloron and $\Zp$  vertices,
we find an identical UV-divergence, the following $s$-dependent terms,
\bea
&& \Lambda_{\Gp} (s, \{m^2_{V_i}\})  = \frac{1}{24} \Lambda_2(s, \mZ)  -\frac{1}{6} \Lambda_2(s, \mG)  \no\\
&& \quad + \frac{3}{2} \Lambda_4(s, \mU, \mU) + \frac{9}{2} \Lambda_4(s, \mG, \mG)~,   \\
&& \Lambda_{\Zp} (s, \{m^2_{V_i}\})   =  \frac{7}{24} \Lambda_2(s, \mZ) 
+ \frac{1}{3} \Lambda_2(s, \mG)  \no\\
&& \quad 
- \frac{3}{4} \Lambda_2(s, \mU)  + 6 \Lambda_4(s, \mU, \mU)  +T_{\rm B-L}^{-1} \left[  \frac{1}{3} \Lambda_2(s, \mG) \right.
\no\\
&& \quad  \left. - \frac{1}{12} \Lambda_2(s, \mZ) - \frac{1}{4} \Lambda_2(s, \mU) \right]~, 
\eea
and the following constant terms
\bea
&& \Lambda_{\Gp}^0 - \Lambda^0_U = -6 +\left( \frac{4 \xG}{\xG-1}  -\frac{9}{2} \right) \log(\xG)   \no \\
&&\qquad + \left(  \frac{2\xZ}{\xZ-1} -\frac{1}{6} \right)\log(\xZ) \,  \stackrel{\rm x_V=1}{\longrightarrow} \,  0~,\qquad \\
&& \Lambda_{\Zp}^0 - \Lambda^0_U  =  -6 +\left( \frac{4 \xG}{\xG-1}  -  \frac{1}{3}   \right)\log(\xG)   \no \\
&&\qquad  + \left(  \frac{2\xZ}{\xZ-1} - \frac{5}{12} \right)\log(\xZ)  \no\\
&& \qquad  +\, T_{\rm B-L}^{-1} \left[ \frac{1}{12} \log(\xZ) - \frac{1}{3} \log(\xG)  \right]  \stackrel{\rm x_V=1}{\longrightarrow} \,  0~,\qquad 
\eea
with $x_V = m_V^2/m_U^2$.
After renormalization,
\be
 \delta V_V(s)_r =
 \delta V_V(s) - \Re\left[ \delta V_U(\mU^2) \right]~.
\ee
In general, both vertex functions $(V=\Gp,\Zp)$  are non-vanishing in the on-shell case
(i.e.~for $s=m^2_V$). However, as shown by the $x_V \to 1$ limits, 
they do vanish on-shell in the $SU(4)$ custodial limit.
 
\subsection{Two-point functions}
The LQ propagator in the Feynman gauge, corrected by resumming one-particle reducible diagrams, can be written as 
\be
\frac{ -i g^{\mu\nu} }{ p^2 - m^2_U } \left[1 + \frac{\alpha_4}{4\pi}  \delta \Sigma_U(p^2) \right] \,,
\ee
where we have already expressed the result in terms of the renormalized mass, and we have taken into account the wave-function renormalization. The non-trival corrections are encoded in the finite term $ \delta \Sigma_U(s)$, that 
we can express in the  on-shell scheme as
\be
 \delta \Sigma_U(s)  = \frac{ \Sigma_U(s) - \Sigma_U(m_U^2)}{s-m_U^2} -  \left. \frac{\partial  \Sigma_U(s)}{\partial s} \right|_{s=m_U^2}\,,
\ee
in terms of the reduced self-energy function $\Sigma_U(s)$. The explicit one loop calculation yields
\bea
&& \Sigma_U(s) = \Sigma^0_U + s  \Sigma^1_U +  \frac{N_f }{3}s\log\left(-\frac{s}{m_U^2}\right)   \no\\
&& +\left[ \frac{m_U^4}{s}\Big(-\frac{\xG^3}{9}-\frac{2\xG^2}{3}+\frac{5\xG}{3}-\frac{8}{9}\Big)   - s\Big(\frac{\xG}{9}-\frac{40}{9}\Big) \right.  \no\\
&& + \left. m_U^2 \Big(\frac{16}{9}+6\xG-\frac{10\xG^2}{9}\Big)\right]F(s, \mU^2, \mG^2)  \no\\
&& +\left[\frac{m_U^4}{s}\Big(-\frac{\xZ^3}{18}-\frac{\xZ^2}{3}+\frac{5\xZ}{6}-\frac{4}{9}\Big) - s\Big( \frac{\xZ}{18}-\frac{20}{9}\Big) \right. \no\\
&&+ \left. m_U^2 \Big(\frac{8}{9}+3\xZ-\frac{5\xZ^2}{9}\Big) \right]F(s, \mU^2, \mZ^2)~. 
\eea
Here $\Sigma^0_U$ and $\Sigma^1_U$ are constant divergent terms, absorbed by the renormalization procedure;
$N_f$ denotes the number of light fermion species transforming in the fundamental  of $SU(4)$;
while $F(s, m_X^2, m_Y^2)$, defined as in~\cite{Bohm:1986rj}, satisfies  $F(0, m_X^2, m_Y^2)=0$. 

The finite correction to the two-point function at $s=0$ assumes the following value 
in the custodial limit
\be
\delta \Sigma_U(0)^{(\rm{cust.})} = 
 \frac{73}{2}-7\sqrt{3}\pi-\frac{N_f}{3}\,  \stackrel{ N_f=4}{\longrightarrow}\,  \approx  - 2.92~.
\ee
Combining the results of 2- and 3-point functions we can evaluate the overall NLO correction
induced by one-particle reducible diagrams to the LQ-mediated four-fermion amplitude at low energies 
\be
\cA^{\rm NLO}_{\rm{1P-red.}} = \cA^{\rm tree} \left\{ 1 + \frac{\alpha_4}{4\pi}  \left[ \delta \Sigma_U(0)+ 2  \delta V_U(0)_r  \right] \right\}~.
\ee
This correction turns out to be very small: in the custodial limit the two terms cancel to a remarkable accuracy, resulting
in a correction below $1\%$ (in absolute size) even for $g_4=3$.  More precisely, setting $g_4=3$ and $N_f=4$,\footnote{The value $N_f=4~(3)$ corresponds 
to the case where we treat the right-handed neutrino mass as light (heavy) compared to $m_U$.} 
the correction lies between -1\% and 0 for $\xG > 0.7$. Sizable positive values can be obtained only 
for small $\xG$ values, but the correction does not exceed 1\% for $\xG> 0.5$.  

\begin{table*}[t]
\begin{center}
\renewcommand{\arraystretch}{1.3}
\begin{tabular}{c||c|c||cccc|c}
 &  \multicolumn{2}{c||}{Tree level}  	 &  \multicolumn{5}{c}{NLO box contributions (in units of $\frac{\alpha_4}{4\pi}$) } \\
\raisebox{5pt}{Operators} &  $U_1$    	&   $\Zp$          &  $[\Gp~U_1]$ & $[\Zp~U_1]$ & $[U_1~U_1]$ &  $[\Zp~\Zp]$ &   Total ($x_V=1$ limit) \\ \hline\hline
$\cO^{U}_{LL} =\frac{1}{2} \big( \cO_{\ell q}^{(1)} + \cO_{\ell q}^{(3)} \big) $  &    1   & 0 	& $\frac{4}{3} f_{\Gp}$ &  $\frac{17}{12} f_{\Zp}$ &  & &    $+\frac{11}{4}$   \\[3pt]  \hline
$\cO^{U}_{LR}  = -2 \cO_{\ell ed q}$       	&   1   & 0 			& $ \frac{16}{3} f_{\Gp}$ & $\frac{23}{12} f_{\Zp} $   &  & &   $+ \frac{29}{4} $\\[3pt]  \hline\hline
$\cO_{\ell q}^{(1)}$ 	&   0      		& $-\frac{1}{4 \xZ }$ &  &  & $-2$  & $- \frac{3}{32 \xZ} $  &   $-\frac{67}{32}$\\[3pt]  \hline
$\cO_{\ell d}$    	&   0  & $-\frac{1}{4 \xZ }$ 	&  & &    $-\frac{1}{2}$  & $\frac{3}{32 \xZ}  $  	&  $-\frac{13}{32}$\\[3pt]  \hline
$\cO_{q e}$       	&  0  & $-\frac{1}{4 \xZ }$ 	&  &  &   $-\frac{1}{2}$  & $\frac{3}{32 \xZ}$  	&  $-\frac{13}{32}$  \\[3pt]  \hline
$\cO_{d e}$       	&  1  & $-\frac{1}{4 \xZ }$ 	&  $\frac{4}{3} f_{\Gp}$ &  $\frac{17}{12} f_{\Zp}$   & $-2$  & $- \frac{3}{32 \xZ} $ 	&  $+\frac{21}{32}$\\[3pt]  \hline
 \end{tabular}
 \caption{Coefficients of the semileptonic operators, normalized as in (\ref{eq:SMEFT}), at tree level and NLO (box contributions only). The NLO 
 results are in units of $\alpha_4/(4\pi)$,  $x_V = m_V^2/m_U^2$ and $f_V = \log(x_V)/(x_V-1)$.
 \label{tab:SMEFT}} 
 \end{center}
\end{table*}

The smallness of this NLO correction can be understood as a consequence of the sudden 
stop in the running of $\alpha_4$ below the LQ mass, when employing a physical (mass-dependent) 
renormalization procedure. The  one-particle reducible diagrams are indeed responsible for the
running of $\alpha_4$ and their combined effect  turns out to be particularly small in the custodial limit, 
where all the heavy particles decouple together, at the scale used to define the 
renormalized coupling.  

We have checked that a similar cancellation holds also for one-particle reducible 
contributions to coloron- and $\Zp$-mediated four-fermion amplitudes. The 
complete expressions for the corresponding self-energy functions, $\Sigma_{\Gp,\Zp}(s)$, 
which coincide with $\Sigma_U(s)$ in the custodial limit, will be reported elsewhere~\cite{Long}.

\subsection{Box diagrams and matching onto the SMEFT}

Due to the effective cancellation of one-particle reducible contributions, 
the only potentially large NLO effects in four-fermion processes originate from box diagrams.

The result of the box diagrams  in the limit of vanishing external momenta 
can be matched onto the basis of dimension-six SMEFT operators~\cite{Grzadkowski:2010es}.
Normalizing the Lagrangian as
\be
\mathcal{L}_{\rm SMEFT}=- \frac{g_4^2}{2 M_U^2} \sum_k\mathcal{C}_k\mathcal{O}_k\,,
\label{eq:SMEFT}
\ee
the Wilson coefficients for the relevant semileptonic operators 
are reported in Table~\ref{tab:SMEFT}. To better illustrate the result,
we perform a change of basis compared to~\cite{Grzadkowski:2010es}
introducing the combinations
\bea
\cO^{U}_{LL}  &=&  (\bar \ell_{L} \gamma_{\mu} q_{L})(\bar q_{L}  \gamma^{\mu}l_{L} )  \no\\
\cO^{U}_{LR}  &=&  - 2 (\bar \ell_{L} e_{R}  ) (\bar  d_{R}  q_{L} )  ~+~{\rm  h.c.}
\eea
which, at the tree level, are the only effective operators 
generated by the LQ exchange. 

Our simplified model features only a single fermion family, hence there is no 
flavor mixing. However, results for realistic models addressing the $B$-physics anomalies 
can be recovered assuming a specific direction for this family in flavor space 
(switching on the Yukawa couplings) and/or introducing appropriate non-trivial
flavor structures in the currents in Eq.~(\ref{eq:U1Lag}), resulting from mixing with heavy fermions.
This way it is easy to realize 
that $\cO^{U}_{LL}$ is the left-handed operator contributing to $b\to c\tau\nu$, which is
present in all the 4321 models, whereas $\cO^{U}_{LR}$ is the scalar operator present in models where the 
$U_1$ has also right-handed couplings~\cite{Bordone:2017bld,Cornella:2019hct}.
As shown in Table~\ref{tab:SMEFT}, in the custodial limit we find a 
$16\%$  ($41\%)$ enhancement for  $\cC^{U}_{LL}$ ($\cC^{U}_{LR}$)
 at NLO, at fixed on-shell coupling $g_4=3$. 
 
 We stress that the effects we have estimated 
are only due to the new dynamics of the heavy vectors, 
therefore they should be considered in addition to the QCD corrections to the high-scale
 matching conditions estimated in~\cite{Aebischer:2018acj}.
 According to this recent analysis, the $\cO(\alpha_s)$ corrections to $\cC^{U}_{LL}$ and $\cC^{U}_{LR}$
 go in the same direction of the  $\cO(\alpha_4)$ ones, i.e.~they enhance the 
coefficients of the effective operators,  and are comparable 
 (significantly smaller) with respect to the $\cO(\alpha_4)$ terms 
 in the case of  $\cC^{U}_{LL}$ ($\cC^{U}_{LR}$).

\begin{figure}[t]
\centering
\includegraphics[width=0.43\textwidth]{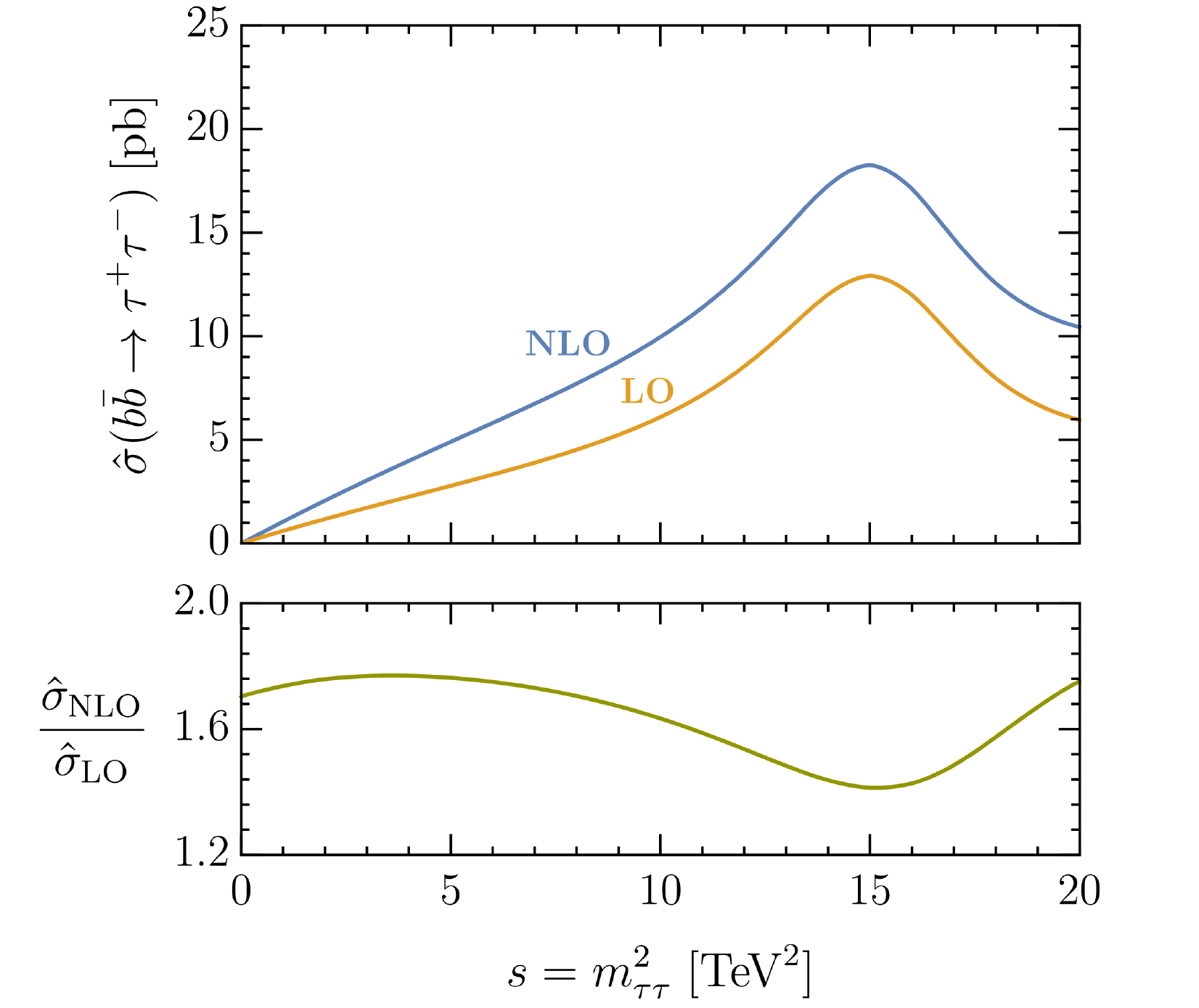}
\caption{\label{fig:sigmaNLO}\textit{Upper panel}:  $b \bar b \to \tau^+ \tau^-$ partonic cross-section at LO and NLO, as a function 
of $s= m^2_{\tau\tau}$.  \textit{Lower panel}: NLO/LO ratio.}
\end{figure}

Our findings have important phenomenological consequences: they imply that all  
collider bounds dominated by the on-shell 
$s$-channel production of the new states (i.e.~the single production of coloron, $Z'$, and leptoquark) 
are significantly weaker at fixed low-energy contribution. This suppression holds
only for the on-shell production of the resonances where: i)~the cross-section can be expressed 
in terms of the on-shell renormalized couplings, ii)~the contribution of the box amplitudes is
subleading (being non-resonant). This is illustrated in Fig.~\ref{fig:sigmaNLO},
where we show the $b \bar b \to \tau \bar \tau$ partonic cross-section at LO and NLO in $\alpha_4$,
within the 4321 model of Ref.~\cite{Bordone:2017bld}, setting $g_4=3$ and $v_1=v_3$ such that $M_V=4$~TeV.
The NLO result takes into account all contributions ($s$-channel $\Zp$, $t$-channel $U_1$, and box amplitudes -- with full kinematical 
dependence). As can be seen, NLO corrections amount to a large increase compared to the LO result. Far from the resonant region, the effect is quite similar to the pure low-energy (SMEFT) regime.
On the other hand, the enhancement becomes smaller close to the $\Zp$ peak, where the process is dominated by the 
on-shell contribution. The suppression is stronger in processes where the resonant amplitude is larger, such as 
$g b \to U_1\tau \to (b \bar \tau) \tau$, dominated by the $U_1$ exchange.

Another important phenomenological  implication of Table~\ref{tab:SMEFT} is the large NLO correction to the 
coefficient of the singlet operator $\cO_{\ell q}^{(1)}$. At tree level, this operator is not generated by $U_1$ 
exchange,\footnote{A potentially dangerous tree-level 
contribution to $\cO_{\ell q}^{(1)}$ is generated by $\Zp$ exchange. However, the flavor-violating component of the latter can be suppressed choosing 
a different flavor-mixing structure for quark-quark and quark-lepton currents~\cite{DiLuzio:2018zxy}. } 
 allowing the model to evade the the strong experimental bounds on $b\to s \nu_\tau \bar\nu_\tau$ and $s \to d \nu_\tau \bar\nu_\tau$
transitions~\cite{Calibbi:2015kma,Barbieri:2015yvd,Buttazzo:2017ixm}. As shown in Table~\ref{tab:SMEFT}, this operator necessarily appears at the one-loop level, even considering only 
box diagrams with leptoquarks. The latter lead to an NLO coefficient for $\cO_{\ell q}^{(1)}$ which is 
$11\%$ of the LO contribution to $\cO^{U}_{LL}$ for $g_4=3$. Currently, this does not pose a serious problem 
for $U_1$ models addressing the $B$-physics anomalies. However, it implies that in $\cB(B \to K^{(*)} \nu \nu)$ and $\cB(K \to \pi \nu \nu)$ 
one should expect $O(10-100\%)$ modifications compared to the corresponding SM predictions 
(see e.g.~\cite{Buttazzo:2017ixm,Bordone:2017lsy}).

\section{Conclusions}
TeV-scale vector leptoquarks are currently the subject of numerous experimental investigations, both at low- and at high-energies. 
If the LQ coupling to SM fermions is large, as expected in motivated models addressing the $B$-physics anomalies,  
potentially large effects beyond tree level should be expected. In this paper we have presented the first estimate of these 
effects in a general class of models based on extensions of the PS gauge symmetry.
As expected, NLO corrections are large, but they are calculable and still within a perturbative regime for $g_4 \lsim 3$.
The main  effect is an enhanced LQ contribution at low-energy, at fixed on-shell couplings. This implies weaker 
constraints from high-energy (on-shell) LQ searches in realistic models addressing $B$-physics anomalies. 

\medskip

%============================================================================
\subsubsection*{Acknowldgements}
%============================================================================
We thank Sandro M\"achler for collaboration in the early stage of this project.
This project has received funding from the European Research Council (ERC) under the European Union's Horizon 2020 research and innovation programme under grant agreement 833280 (FLAY), and by the Swiss National Science Foundation (SNF) under contract 200021-159720. The work of J.F. was also supported in part by the Generalitat Valenciana under contract SEJI/2018/033.

%%%%%%%%%%%%%%%%%%%%%%%%%%%%%%%%%%%%%%%%%%%%%%%%%%

\end{document}